\documentclass[useAMS,usenatbib]{mn2e}

\usepackage{graphicx}
\usepackage{epsfig,floatflt}
\usepackage{times}

\title[Spectroscopic binaries among bright Cepheids]
{Discovery of the spectroscopic binary nature 
of three bright southern Cepheids}

\author[Szabados et al.]{L. Szabados$^{1}$, R.~I. Anderson$^2$, 
A. Derekas$^{1,3}$, 
L. L. Kiss$^{1,3,4}$, T. Szalai$^5$,  
\newauthor  P. Sz\'ekely$^6$, J.~L. Christiansen$^7$
\\$^1$Konkoly Observatory, Research Centre for Astronomy and Earth 
Sciences, Hungarian Academy of Sciences, Konkoly Thege Mikl\'os \'ut 15-17,\\
H-1121 Budapest, Hungary\\
$^2$Observatoire de Gen\`eve, Universit\'e de Gen\`eve, 
51 Ch. des Maillettes, CH-1290 Versoix, Switzerland\\
$^3$Sydney Institute for Astronomy, School of Physics, 
University of Sydney, NSW 2006, Australia\\
$^4$ELTE Gothard-Lend\"ulet Research Group, Szent Imre herceg \'ut 112,
H-9700 Szombathely, Hungary\\
$^5$Department of Optics and Quantum Electronics, 
University of Szeged, D\'om t\'er 9, H-6720 Szeged, Hungary\\
$^6$Department of Experimental Physics, University of Szeged, 
Szeged H-6720, Hungary\\
$^7$SETI Institute/NASA Ames Research Center, M/S 244-30, 
Moffett Field, CA 94035, USA
}

\begin{document}

\date{Accepted Received ; in original form }

\pagerange{\pageref{firstpage}--\pageref{lastpage}} \pubyear{2013}

\maketitle

\label{firstpage}

\begin{abstract}
We present an analysis of spectroscopic radial velocity and 
photometric data of three bright Galactic Cepheids: LR~Trianguli 
Australis (LR~TrA), RZ~Velorum (RZ~Vel), and BG~Velorum (BG~Vel). 
Based on new radial velocity data, these Cepheids have been found 
to be members of spectroscopic binary systems. 

The ratio of the peak-to-peak radial velocity amplitude 
to photometric amplitude indicates the presence of a companion 
for LR~TrA and BG~Vel. {\it IUE} spectra indicate that 
the companions of RZ~Vel and BG~Vel cannot be hot stars.

The analysis of all available photometric data revealed that the 
pulsation period of RZ~Vel and BG~Vel varies monotonically, due to 
stellar evolution. Moreover, the longest period Cepheid in this sample, 
RZ~Vel, shows period fluctuations superimposed on the monotonic period 
increase. The light-time effect interpretation of the observed pattern
needs long-term photometric monitoring of this Cepheid. The pulsation 
period of LR~TrA has remained constant since the discovery of its 
brightness variation.

Using statistical data, it is also shown that a large number of
spectroscopic binaries still remain to be discovered among bright 
classical Cepheids.

\end{abstract}

\begin{keywords}
binaries: spectroscopic -- stars: variables: Cepheids
\end{keywords}

\section{Introduction}
\label{intro}

Classical Cepheid variable stars are primary distance 
indicators and rank among standard candles for establishing 
the cosmic distance scale, owing to the famous period-luminosity 
($P$--$L$) relationship.
Companions to Cepheids, however, complicate the situation.
The contribution of the secondary star to the observed 
brightness has to be taken into account when involving any
particular Cepheid in the calibration of the $P$--$L$ relationship.

Binaries among Cepheids are not rare at all: their frequency 
exceeds 50 per cent for the brightest Cepheids, while among the 
fainter Cepheids an observational selection effect encumbers 
revealing binarity \citep{Sz03a}.

\begin{figure}
\includegraphics[height=48mm, angle=0]{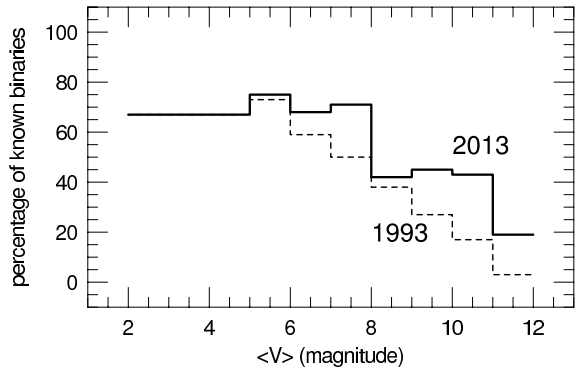}
\caption{Percentage of known binaries among Galactic
classical Cepheids as a function of the mean apparent
visual brightness in 1993 and 2013. The decreasing influence
of the observational selection effect is noticeable.}
\label{fig-comparison}
\end{figure}

\begin{figure}
\vspace*{7mm}
\includegraphics[height=48mm, angle=0]{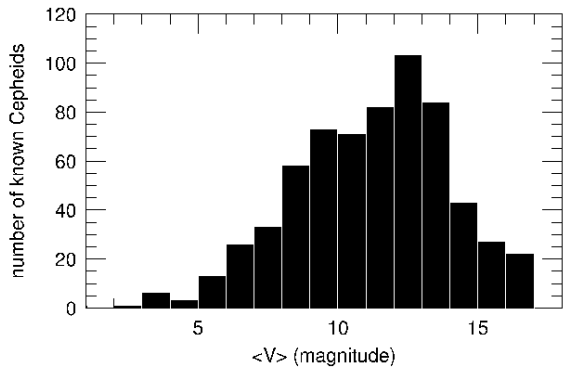}
\caption{Histogram showing the number distribution of known
Galactic classical Cepheids as a function of their mean 
apparent visual brightness.}
\label{fig-histogram}
\end{figure}

Owing to some observational projects aimed at obtaining new
radial velocities (RVs) of numerous Cepheids carried out during
the last decades, a part of the selection effect has been
removed. This progress is visualized in Fig.~\ref{fig-comparison}
where the current situation is compared with that 20 years ago.
The data have been taken from the on-line data base on binaries 
among Galactic Cepheids (http://www.konkoly.hu/CEP/orbit.html).
To get rid of the fluctuation at the left-hand part of the diagram,
brightest Cepheids ($\langle V \rangle <5$~mag) were merged in a
single bin because such stars are extremely rare among Cepheids
-- see the histogram in Fig.~\ref{fig-histogram}.

In the case of pulsating variables, like Cepheids, spectroscopic
binarity manifests itself in a periodic variation of the
$\gamma$-velocity (i.e., the RV of the mass centre of the Cepheid). 
In practice, the orbital RV variation of the Cepheid component is 
superimposed on the RV variations of pulsational origin. 
To separate orbital and pulsational effects, knowledge of the 
accurate pulsation period is essential, especially when comparing 
RV data obtained at widely differing epochs. Therefore, the pulsation
period and its variations have been determined with the method of
the O$-$C diagram \citep{S05} for each target Cepheid. Use of the
accurate pulsation period obtained from the photometric data is a 
guarantee for the correct phase matching of the (usually less
precise) RV data.

In this paper we point out spectroscopic binarity of three 
bright Galactic Cepheids by analysing RV data. The structure of 
this paper is as follows. The new observations and the equipment 
utilized are described in Sect.~\ref{newdata}. Section~\ref{results} 
is devoted to the results on the three new spectroscopic binary (SB) 
Cepheids: LR~Trianguli Australis, RZ~Velorum, and BG~Velorum. 
Basic information on these Cepheids is given in Table~\ref{obsprop}. 
Finally, Section~\ref{concl} contains our conclusions.

\begin{table}  
\begin{center}  
\caption{Basic data of the programme stars 
and the number of spectra.} 
\label{obsprop}  
\begin{tabular}{|lccccc|} 
\hline  
Cepheid & $\langle V \rangle$ & $P$ & Mode & \multicolumn{2}{c}{Number of spectra}\\
& (mag) & (d)& of pulsation & SSO & CORALIE \\
\hline  
LR~TrA &  7.80 &  2.428289 & first overtone & 10 & 32\\ 
RZ~Vel &  7.13 & 20.398532 & fundamental    & 30 & 67\\
BG~Vel &  7.69 &  6.923843 & fundamental    & 27 & 33\\  
\hline   
\end{tabular} 
\end{center}  
\end{table}


\section{New observations}
\label{newdata}

\subsection{Spectra from the Siding Spring Observatory}
\label{SSO}

We performed an RV survey of Cepheids with the 2.3~m ANU 
telescope located at the Siding Spring Observatory (SSO), 
Australia. The main aim of the project was to detect Cepheids 
in binary systems by measuring changes in the mean values of 
their RV curve which can be interpreted as the orbital 
motion of the Cepheid around the centre-of-mass in a binary 
system (change of $\gamma$-velocity). The target list was 
compiled to include Cepheids with a single-epoch RV phase curve 
or without any published RV data. Several Cepheids suspected 
to be members of SB systems were also put on the target list. 
In 64 nights between 2004 October and 2006 March we monitored 
40 Cepheids with pulsation periods between 2 and 30 d.
Additional spectra of some targets were obtained in 2007 August.

Medium-resolution spectra were taken with the Double Beam 
Spectrograph using the 1200~mm$^{-1}$ gratings in both arms of 
the spectrograph. The projected slit width was 2 arcsec 
on the sky, which was about the median seeing during our 
observations. The spectra covered the wavelength ranges 
4200--5200~\AA\ in the blue arm and 5700--6700~\AA\ in the red 
arm. The dispersion was 0.55~\AA~pixel$^{-1}$, leading to a nominal 
resolution of about 1~\AA.

All spectra were reduced with standard tasks in {\sc iraf}
\footnote{{\sc iraf} is distributed by the National Optical 
Astronomy Observatories, which are operated by the Association
of Universities for Research in Astronomy, Inc., under 
cooperative agreement with the National Science Foundation.}.
Reduction consisted of bias and flat-field corrections, 
aperture extraction, wavelength calibration, and continuum 
normalization. We checked the consistency of wavelength 
calibrations via the constant positions of strong telluric 
features, which proved the stability of the system. 
RVs were determined only for the red arm data 
with the task {\it fxcor\/}, applying the cross-correlation 
method using a well-matching theoretical template spectrum 
from the extensive spectral library of \citet{Metal05}. Then, 
we made barycentric corrections to every single RV value. 
This method resulted in a 1-2~km~s$^{-1}$ uncertainty in the 
individual RVs, while further tests have shown that our 
absolute velocity frame was stable to within 
$\pm$2--3~km~s$^{-1}$. This level of precision is sufficient
to detect a number of Cepheid companions, as they can often
cause $\gamma$-velocity changes well above 10~km~s$^{-1}$.

Discovery of six SBs among the 40 target Cepheids was 
already reported by \citet{Szetal13}. The binarity of the 
three Cepheids announced here could be revealed by involving 
independently obtained additional data (see Section~\ref{coralie}).
The individual RV data of the rest of the Cepheid targets
will be published together with the results of the analysis
of the spectra.

\subsection{CORALIE observations from La Silla}
\label{coralie}

All three Cepheids were among the targets during multiple observing 
campaigns between 2011 April and 2012 May using the fibre-fed 
high-resolution ($R \sim 60000$) echelle spectrograph 
\textit{CORALIE} mounted on the Swiss 1.2\,m Euler telescope at 
ESO La Silla Observatory, Chile. The instrument's design is 
described in \citet{Qetal01}; recent instrumental updates 
can be found in \citet{Setal10}. 

When it turned out that these three Cepheids have variable
$\gamma$-velocities, several new spectra were obtained in
2012 December - 2013 January and 2013 April.

The spectra are reduced by the efficient online reduction 
pipeline that performs bias correction, cosmics removal, 
and flatfielding using tungsten lamps. ThAr lamps are used 
for the wavelength calibration. The reduction pipeline directly 
determines the RV via cross-correlation \citep{Betal96} 
using a mask that resembles a G2 spectral type. 
The RV stability of the instrument is excellent and for 
non-pulsating stars the RV precision is limited by photon noise;
(see e.g., \citet{Petal02}). However, the precision achieved for 
Cepheids is lower due to line asymmetries. We estimate a typical 
precision of $\sim$ 0.1\,km\,s$^{-1}$ (including systematics due
to pulsation) per data point for our data.

\section{Results for individual Cepheids}
\label{results}

\subsection{LR~Trianguli Australis}
\label{lrtra}

\paragraph*{Accurate value of the pulsation period}
\label{lrtra-period}

The brightness variability of LR~TrA (HD\,137626, $\langle V \rangle
= 7.80$\,mag) was revealed by \citet{Setal66} based on the Bamberg
photographic patrol plates. The Cepheid nature of variability and the
first values of the pulsation period was determined by \citet{E83}. 
This Cepheid pulsates in the first-overtone mode; therefore, it has 
a small pulsational amplitude and nearly-sinusoidal light and
velocity curves. 

In the case of Cepheids pulsating with a low amplitude, the O$-$C 
diagram constructed for the median brightness (the mid-point 
between the faintest and the brightest states) is more reliable 
than that based on the moments of photometric maxima \citep{Detal12}. 
Therefore we determined the accurate value of the pulsation period 
by constructing an O$-$C diagram for the moments of median brightness 
on the ascending branch of the light curve since this is the phase when 
the brightness variations are steepest during the whole pulsational 
cycle.

All published photometric observations of LR~TrA covering three 
decades were re-analysed in a homogeneous manner to determine 
seasonal moments of the chosen light-curve feature. The relevant data 
listed in Table~\ref{tab-lrtra-oc} are as follows:\\
Column~1: heliocentric moment of the selected light-curve feature
(median brightness on the ascending branch for LR~TrA, maximum
brightness for both RZ~Vel and BG~Vel, see Tables~\ref{tab-rzvel-oc}
and \ref{tab-bgvel-oc}, respectively;\\
Col.~2: epoch number, $E$, as calculated from Equation~(\ref{lrtra-ephemeris}):
\vspace{-1mm}
\begin{equation}
C = 2\,453\,104.9265 + 2.428\,289{\times}E 
\label{lrtra-ephemeris}
\end{equation}
\vspace{-3mm}
$\phantom{mmmmm}\pm0.0037\phantom{}\pm0.000\,003$

\noindent (this ephemeris has been obtained by the weighted 
least squares parabolic fit to the O$-$C differences);\\
\noindent Col.~3: the corresponding O$-$C value;\\
Col.~4: weight assigned to the O$-$C value (1, 2, or 3 
depending on the quality of the light curve leading to 
the given difference);\\
Col.~5: reference to the origin of data.\\

\begin{figure}
\includegraphics[height=44mm, angle=0]{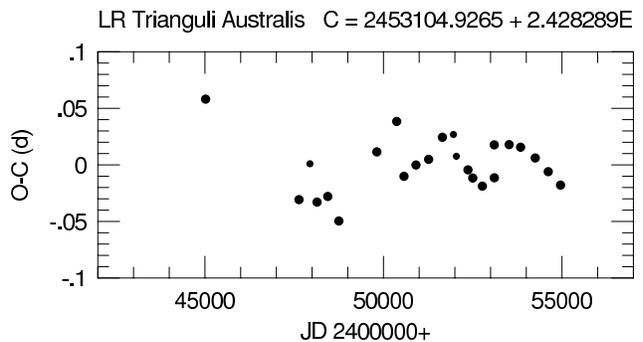}
\caption{O$-$C diagram of LR~TrA. The plot can be
approximated by a constant period.}
\label{fig-lrtra-oc}
\end{figure}

The O$-$C diagram of LR~TrA based on the O$-$C values listed
in Table~\ref{tab-lrtra-oc} is plotted in 
Fig.~\ref{fig-lrtra-oc}. The plot can be approximated by a 
constant period by the ephemeris (\ref{lrtra-ephemeris}) for 
the moments of median brightness on the ascending branch. The 
scatter of the points in Fig.~\ref{fig-lrtra-oc} reflects the 
observational error and uncertainties in the analysis of the data.

\begin{table}
\caption{O$-$C values of LR~TrA (see the 
description in Sect.~\ref{lrtra-period}).}
\begin{tabular}{l@{\hskip2mm}r@{\hskip2mm}r@{\hskip2mm}c@{\hskip2mm}l}
\hline
\noalign{\vskip 0.2mm}
JD$_{\odot}$ & $E\ $ & O$-$C & $W$ & Data source\\
2\,400\,000 + &&&\\
\noalign{\vskip 0.2mm}
\hline
\noalign{\vskip 0.2mm}
45018.7822 & $-$3330&   0.0581  &3 & \citet{E83}\\
47633.9607 & $-$2253& $-$0.0307 & 3 & \citet{Aetal90}\\
47939.9568 & $-$2127&  0.0010  &2 & {\it Hipparcos} \citep{ESA97}\\
48139.0426 & $-$2045& $-$0.0329 & 3&  {\it Hipparcos} \citep{ESA97}\\
48440.1554 & $-$1921& $-$0.0279 & 3& {\it Hipparcos} \citep{ESA97}\\
48750.9547 & $-$1793& $-$0.0496 & 3& {\it Hipparcos} \citep{ESA97}\\
49814.6064 & $-$1355&   0.0115 & 3 & \citet{B08}\\
50370.7115 & $-$1126&   0.0384 & 3 & \citet{B08}\\
50574.6393 & $-$1042& $-$0.0101 & 3& \citet{B08}\\
50909.7531 & $-$904& $-$0.0001 & 3 & \citet{B08}\\
51264.2883 & $-$758&   0.0049 & 3 & \citet{B08}\\
51650.4058 & $-$599&   0.0244 & 3 & \citet{B08}\\
51958.8010 & $-$472&   0.0269 & 2 & \citet{B08}\\
52041.3435 & $-$438&   0.0076 & 2 & ASAS \citep{P02}\\
52366.7222 & $-$304& $-$0.0044 & 3 & \citet{B08}\\
52500.2709 & $-$249& $-$0.0116 & 3 & ASAS \citep{P02}\\
52769.8038 & $-$138& $-$0.0188 & 3 & ASAS \citep{P02}\\
53102.5159 &  $-$1&   0.0177 & 3 & \citet{B08}\\
53104.9151 &     0& $-$0.0114 & 3&  ASAS \citep{P02}\\
53520.1818 &   171&   0.0179 & 3 & ASAS \citep{P02}\\
53840.7137 &   303&   0.0156 & 3 & ASAS \citep{P02}\\
54251.0850 &   472&   0.0061 & 3& ASAS \citep{P02}\\
54615.3163 &   622& $-$0.0060&  3 & ASAS \citep{P02}\\
54960.1214 &   764& $-$0.0179&  3 & ASAS \citep{P02}\\
\noalign{\vskip 0.2mm}
\hline
\end{tabular}
\label{tab-lrtra-oc}
\end{table}

\paragraph*{Binarity of LR~TrA}
\label{lrtra-bin}

\begin{figure}
\includegraphics[height=48mm, angle=0]{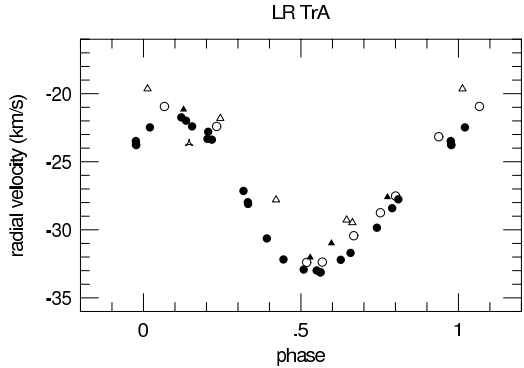}
\caption{Merged RV phase curve of LR~TrA. The different symbols
mean data from different years: 2005: filled triangles; 2006: 
empty triangles; 2007: triangular star; 2012: filled circles; 
2013: empty circles. The zero phase was arbitrarily chosen at 
JD\,2\,400\,000.0 (in all phase curves in this paper).}
\label{fig-lrtra-vrad}
\end{figure}

\begin{figure}
\includegraphics[height=40mm, angle=0]{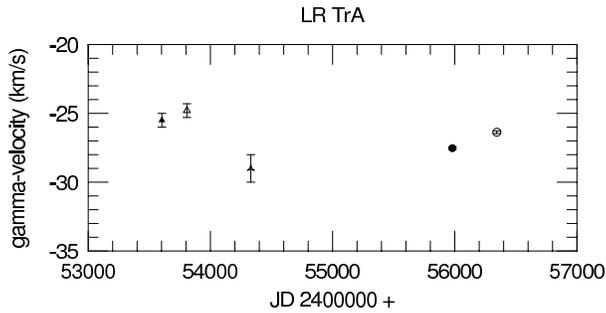}
\caption{Temporal variation in the $\gamma$-velocity of LR~TrA.
The symbols for the different data sets are the same as
in Fig.~\ref{fig-lrtra-vrad}.}
\label{fig-lrtra-vgamma}
\end{figure}

\begin{table}
\caption{RV values of LR TrA from the SSO spectra. 
This is only a portion of the full version available online as Supporting
Information.}
\begin{tabular}{lr}
\hline
\noalign{\vskip 0.2mm}
JD$_{\odot}$ & $v_{\rm rad}$  \ \\
2\,400\,000 + &(km\,s$^{-1}$)\\
\noalign{\vskip 0.2mm}
\hline
\noalign{\vskip 0.2mm}
53599.9325 &$-$21.2\\
53600.9086 &$-$32.0\\
53603.9327 &$-$27.6\\
53605.9290 &$-$31.0\\
53805.1657 &$-$29.3\\
\noalign{\vskip 0.2mm}
\hline
\end{tabular}
\label{tab-lrtra-data}
\end{table}

\begin{table}
\caption{CORALIE velocities of LR TrA. 
This is only a portion of the full version 
available online as Supporting Information.}
\begin{tabular}{lrc}
\hline
\noalign{\vskip 0.2mm}
JD$_{\odot}$ & $v_{\rm rad}$ \ & $\sigma$ \\
2\,400\,000 + &(km\,s$^{-1}$) & (km\,s$^{-1}$)\\
\noalign{\vskip 0.2mm}
\hline
\noalign{\vskip 0.2mm}
55938.8701 & $-$27.97 & 0.05\\
55938.8718 & $-$28.10 & 0.05\\  
55939.8651 & $-$29.85 & 0.02\\  
55940.8686 & $-$22.40 & 0.03\\  
55941.8579 & $-$33.14 & 0.04\\  
\noalign{\vskip 0.2mm}
\hline
\end{tabular}
\label{tab-lrtra-coralie-data}
\end{table}

\begin{table}
\caption{$\gamma$-velocities of LR~TrA.}
\begin{tabular}{lccl}
\hline
\noalign{\vskip 0.2mm}
Mid-JD & $v_{\gamma}$ & $\sigma$ & Data source \\
2\,400\,000+ & (km\,s$^{-1}$)& (km\,s$^{-1}$) & \\
\noalign{\vskip 0.2mm}
\hline
\noalign{\vskip 0.2mm}
53603 & $-$25.5 & 0.5 & Present paper\\
53808 & $-$24.8 & 0.5 & Present paper\\
54331 & $-$29.0 & 1.0 & Present paper\\
55981 & $-$27.5 & 0.1 & Present paper\\
56344 & $-$26.4 & 0.1 & Present paper\\
\noalign{\vskip 0.2mm}
\hline
\end{tabular}
\label{tab-lrtra-vgamma}
\end{table}

There are no earlier RV data on this bright Cepheid. Our new data 
listed in Tables~\ref{tab-lrtra-data} and \ref{tab-lrtra-coralie-data} 
have been folded on the accurate pulsation period given in the
ephemeris (see Equation~\ref{lrtra-ephemeris}). The merged RV phase 
curve is plotted in Fig.~\ref{fig-lrtra-vrad}. Both individual 
data series could be split into seasonal subsets.

Variability in the $\gamma$-velocity is obvious. The 
$\gamma$-velocities (together with their uncertainties) are 
listed in Table~\ref{tab-lrtra-vgamma}. The $\gamma$-velocity in
2007 is more uncertain than in other years because this value
is based on a single spectrum. Systematic errors can be excluded. 
Dozens of Cepheids in our sample with non-varying
$\gamma$-velocities indicate stability of the equipment and 
reliability of the data reduction. Fig.~\ref{fig-lrtra-vgamma} 
is a better visualization of the temporal variation in the 
$\gamma$-velocity. The seasonal drift in the $\gamma$-velocity 
is compatible with both short and long orbital periods.

The photometric contribution of the companion star decreases
the observable amplitude of the brightness variability as
deduced from the enhanced value of the ratio of the RV and
photometric amplitudes \citep{KSz09}. This is an additional
(although slight) indication of binarity of LR~TrA.

\subsection{RZ~Velorum}
\label{rzvel}

\paragraph*{Accurate value of the pulsation period}
\label{rzvel-period}

The brightness variability of RZ~Vel (HD\,73502, $\langle V \rangle
= 7.13$\,mag) was revealed by Cannon \citep{P09}. The Cepheid
nature of variability and the pulsation period were established by 
\citet{H36} based on the Harvard and Johannesburg photographic plate 
collection which was further investigated by \citet{Oo36}.

This is the longest period Cepheid announced in this paper and it has 
been frequently observed from the 1950s, first photoelectrically, 
then in the last decades by CCD photometry. The photometric coverage 
of RZ~Vel was almost continuous in the last 20 years thanks to 
observational campaigns by \citet{B08} and his co-workers, as well as
the ASAS photometry \citep{P02}.

Long-period Cepheids are usually fundamental pulsators and they 
oscillate with a large amplitude resulting in a light curve with
sharp maximum.

The O$-$C diagram of RZ~Vel was constructed for the moments of 
maximum brightness based on the photoelectric and CCD photometric 
data (see Table~\ref{tab-rzvel-oc}). The weighted least squares 
parabolic fit to the O$-$C values resulted in the ephemeris:
\vspace{-1mm}
\begin{equation}
C = 2\,442\,453.6630 + 20.398\,532{\times}E + 1.397\times 10^{-6} E^2
\label{rzvel-ephemeris}
\end{equation}
\vspace{-3mm}
$\phantom{mmmmm}\pm0.0263\phantom{l}\pm 0.000\,080 \phantom{mm}
\pm 0.191\times 10^{-6}$

\begin{table}
\caption{O$-$C values of RZ~Vel (description of the columns
is given in Sect.~\ref{lrtra-period}).}
\begin{tabular}{l@{\hskip2mm}r@{\hskip2mm}r@{\hskip2mm}c@{\hskip2mm}l}
\hline
\noalign{\vskip 0.2mm}
JD$_{\odot}$ & $E\ $ & O$-$C & $W$ & Data source\\
2\,400\,000 + &&&\\
\noalign{\vskip 0.2mm}
\hline
\noalign{\vskip 0.2mm}
33784.5646 &$-$425 &  0.2777 & 1 & \citet{Eetal57}\\
34804.5174 &$-$375 &  0.3039 & 1 & \citet{Wetal58}\\
34845.2119 &$-$373 &  0.2013 & 3 & \citet{Eetal57}\\
35192.0024 &$-$356 &  0.2168 & 1 & \citet{I61}\\
40760.8647  &$-$83 &  0.2799 & 3 & \citet{P76}\\
41719.0924  &$-$36 &$-$0.2234&  3& \citet{M75}\\
41862.1249  &$-$29 &  0.0193 & 3 & \citet{Detal77}\\
42453.6330 &    0 &$-$0.0030 & 3 & \citet{Detal77}\\
44371.0472 &   94 &$-$0.0778 & 3 & \citet{CC85}\\
44391.3842 &   95 &$-$0.1393 & 2 & \citet{E82}\\
45003.2906 &  125 &$-$0.1889 & 3 & \citet{CC85}\\
48226.4369 &  283 &$-$0.0107 & 3 & {\it Hipparcos} \citep{ESA97}\\
48797.5877 &  311 &$-$0.0188 & 3 & {\it Hipparcos} \citep{ESA97}\\
49185.1653 &  330 &$-$0.0133 & 1 & Walker \& Williams (unpublished)\\
49817.8011 &  361 &  0.2680  &3 & \citet{B08}\\
50144.1979 &  377 &  0.2883 & 2 & \citet{B02}\\
50389.0443 &  389 &  0.3524 & 3 & \citet{B08}\\
50511.3662 &  395 &  0.2831 & 3 & \citet{B02}\\
50572.4468 &  398 &  0.1681 & 3 & \citet{B08}\\
50899.0581 &  414 &  0.4029 & 3 & \citet{B08}\\
51266.1488 &  432 &  0.3200 & 3 & \citet{B08}\\
51653.7650 &  451 &  0.3641 & 3 & \citet{B08}\\
51939.2846 &  465 &  0.3042 & 2 & ASAS \citep{P02}\\
51959.7692 &  466 &  0.3903 & 3 & \citet{B08}\\
52347.4262 &  485 &  0.4752 & 3 & \citet{B08}\\
52653.3896 &  500 &  0.4606 & 3 & ASAS \citep{P02}\\
52653.4100 &  500 &  0.4810 & 3 & \citet{B08}\\
53000.1794 &  517 &  0.4754 & 3 & ASAS \citep{P02}\\
53000.2610 &  517 &  0.5570 & 3 & \citet{B08}\\
53428.4384 &  538 &  0.3652 & 3 & ASAS \citep{P02}\\
53754.8864 &  554 &  0.4367 & 3 & ASAS \citep{P02}\\   
54183.1657 &  575 &  0.3468 & 3 & ASAS \citep{P02}\\
54509.5729 &  591 &  0.3775 & 3 & ASAS \citep{P02}\\
54815.4343 &  606 &  0.2609 & 3 & ASAS \citep{P02}\\
55121.3569 &  621 &  0.2055 & 2 & ASAS \citep{P02}\\
\noalign{\vskip 0.2mm}
\hline
\end{tabular}
\label{tab-rzvel-oc}
\end{table}

\begin{figure}
\includegraphics[height=55mm, angle=0]{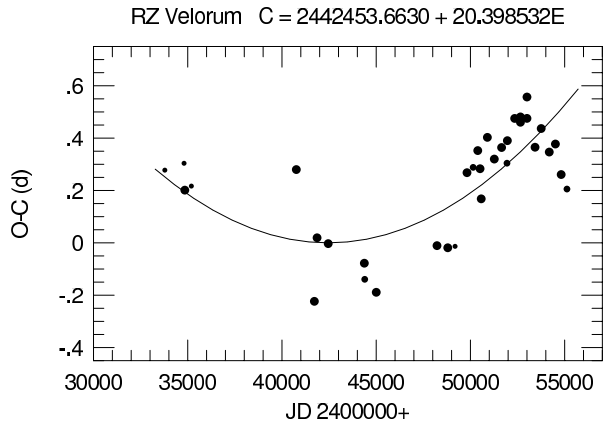}
\caption{O$-$C diagram of RZ~Vel. The plot can be
approximated by a parabola indicating a continuously
increasing period.}
\label{fig-rzvel-oc}
\end{figure}

The O$-$C diagram of RZ~Vel plotted in Fig.~\ref{fig-rzvel-oc} 
indicates a continuously increasing pulsation period with a period 
jitter superimposed. This secular period increase has been caused 
by stellar evolution: while the Cepheid crosses the instability 
region towards lower temperatures in the Hertzsprung--Russell 
diagram, its pulsation period is increasing. 
Continuous period variations (of either sign) often occur in 
the pulsation of long-period Cepheids \citep{Sz83}.

\begin{figure}
\includegraphics[height=44mm, angle=0]{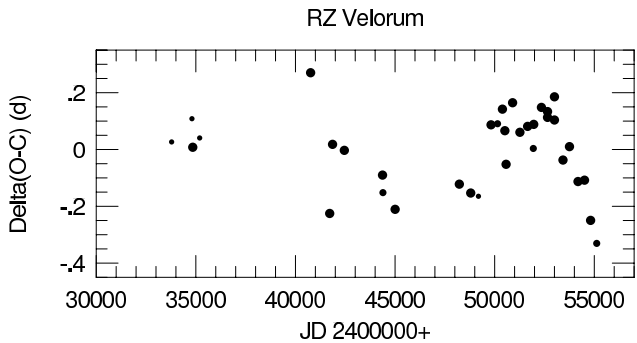}
\caption{$\Delta(O-C)$ diagram of RZ~Vel.}
\label{fig-rzvel-oc2}
\end{figure}

Fig.~\ref{fig-rzvel-oc2} shows the O$-$C residuals after 
subtracting the parabolic fit defined by 
Equation~(\ref{rzvel-ephemeris}). If the wave-like fluctuation seen in 
this $\Delta (O-C)$ diagram turns out to be periodic, it would
correspond to a light-time effect in a binary system. In line with 
the recent shortening in the pulsation period, the current value 
of the pulsation period is $20.396671 \pm 0.000200$ days (after 
JD~2\,452\,300). 

\paragraph*{Binarity of RZ~Vel}
\label{rzvel-bin}

\begin{figure}
\includegraphics[height=55mm, angle=0]{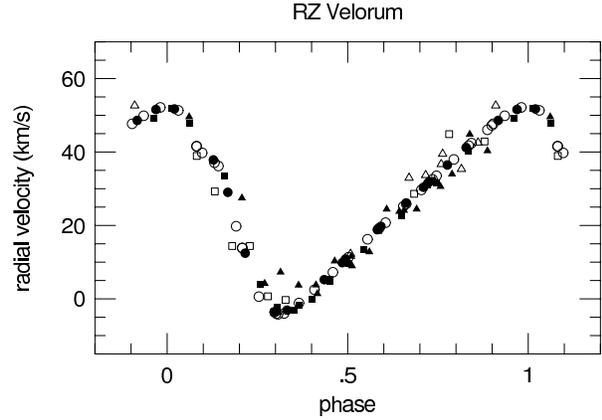}
\caption{RV phase curve of RZ~Vel. Data obtained
between 1996 and 2013 are included in this plot. The meaning
of various symbols is explained in the text.}
\label{fig-rzvel-vrad}
\end{figure}

\begin{figure}
\includegraphics[height=42mm, angle=0]{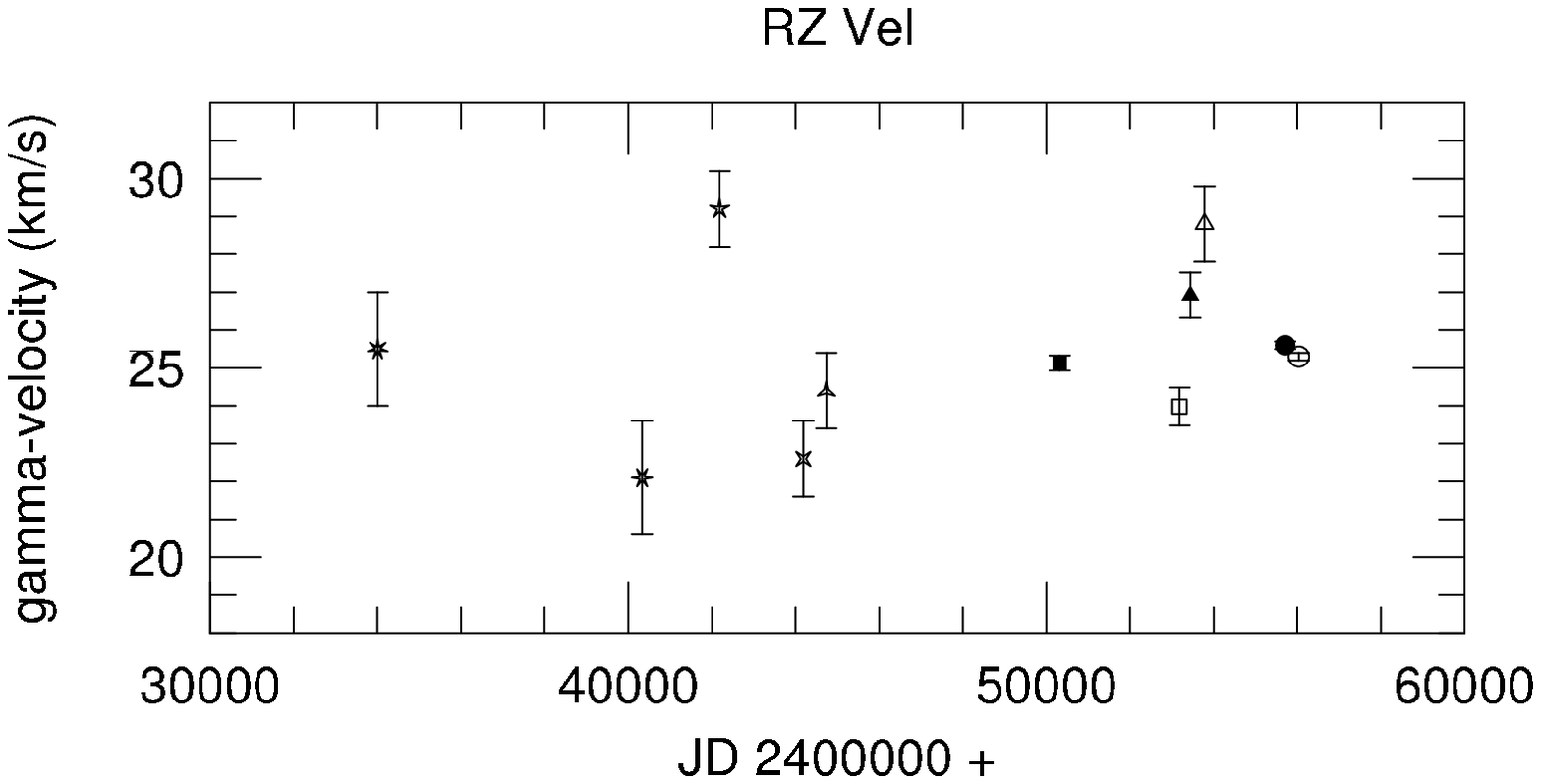}
\caption{$\gamma$-velocities of RZ~Velorum. The symbols for 
the different data sets are the same as in 
Fig.~\ref{fig-rzvel-vrad}.}
\label{fig-rzvel-vgamma}
\end{figure}

\begin{table}
\caption{RV values of RZ Vel from the SSO spectra.
(This is only a portion of the full version 
available online as Supporting
Information.)}
\begin{tabular}{lr}
\hline
\noalign{\vskip 0.2mm}
JD$_{\odot}$ & $v_{\rm rad}$ \  \\
2\,400\,000 + &(km\,s$^{-1}$)\\
\noalign{\vskip 0.2mm}
\hline
\noalign{\vskip 0.2mm}
53307.2698 &4.2\\
53310.2504 &1.4\\
53312.2073 &9.0\\
53364.2062 &49.6\\
53367.1823 &27.5\\
\noalign{\vskip 0.2mm}
\hline
\end{tabular}
\label{tab-rzvel-data}
\end{table}

\begin{table}
\caption{CORALIE velocities of RZ Vel.
(This is only a portion of the full version available 
online as Supporting Information.)}
\begin{tabular}{lrc}
\hline
\noalign{\vskip 0.2mm}
JD$_{\odot}$ & $v_{\rm rad}$ \ & $\sigma$ \\
2\,400\,000 + &(km\,s$^{-1}$) & (km\,s$^{-1}$)\\
\noalign{\vskip 0.2mm}
\hline
\noalign{\vskip 0.2mm}
55654.5528 & $-$3.08 & 0.02\\
55656.6626 &  5.23 & 0.01\\  
55657.6721 &  9.86 & 0.02\\  
55659.6585 & 18.85 & 0.03\\  
55662.5137 & 31.50 & 0.01\\  
\noalign{\vskip 0.2mm}
\hline
\end{tabular}
\label{tab-rzvel-coralie-data}
\end{table}

\begin{table}
\caption{$\gamma$-velocities of RZ~Vel.}
\begin{tabular}{lccl}
\hline
\noalign{\vskip 0.2mm}
Mid-JD & $v_{\gamma}$ & $\sigma$ & Data source \\
2\,400\,000+ & (km\,s$^{-1}$)& (km\,s$^{-1}$) & \\
\noalign{\vskip 0.2mm}
\hline
\noalign{\vskip 0.2mm}
34009  &25.5 &1.5& \citet{S55}\\
40328  &22.1 &1.5& \citet{LE68,LE80}\\
42186  &29.2 &1.0& \citet{CC85}\\
44186  &22.6 &1.0& \citet{CC85}\\
44736  &24.4 &1.0& \citet{CC85}\\
50317  &25.1 &0.2& \citet{B02}\\
53184  &24.0 &0.5& \citet{Netal06}\\
53444  &26.9 &0.6& Present paper\\
53783  &28.8 &1.0& Present paper\\
55709  &25.6 &0.1& Present paper\\
56038  &25.3 &0.1& Present paper\\
\noalign{\vskip 0.2mm}
\hline
\end{tabular}
\label{tab-rzvel-vgamma}
\end{table}

There are several data sets of RV observations available in
the literature for RZ~Vel: those published by \citet{S55}, 
\citet{LE68,LE80}, \citet{CC85}, \citet{B02}, and 
\citet{Netal06}. Our individual RV data are listed in 
Tables~\ref{tab-rzvel-data} and \ref{tab-rzvel-coralie-data}.

Based on these data, the RV phase curve has been constructed 
using the 20.398532~d pulsation period appearing in 
Equation~(\ref{rzvel-ephemeris}). In view of the complicated pattern 
of the O$-$C diagram the RV data have been folded on by taking 
into account the proper phase correction for different data 
series. The merged RV phase curve is plotted in 
Fig.~\ref{fig-rzvel-vrad}. For the sake of clarity, RV data 
obtained before JD\,2\,450\,000 have not been plotted here 
because of the wider scatter of these early RV data but the 
$\gamma$-velocities were determined for each data set. The 
individual data series are denoted by different symbols: 
filled squares mean data by \citet{B02}, empty squares those 
by \citet{Netal06}, and our 2005, 2006, 2012 and 2013 data are 
denoted by filled triangles, empty triangles, filled circles and 
empty circles, respectively. The wide scatter in this merged RV 
phase curve plotted in Fig.~\ref{fig-rzvel-vrad} is due to a variable 
$\gamma$-velocity. 

The $\gamma$-velocities determined from each data set (including 
the earlier ones) are listed in Table~\ref{tab-rzvel-vgamma} and 
are plotted in Fig.~\ref{fig-rzvel-vgamma}. The plot implies
that RZ~Vel is really an SB as suspected by \citet{B02} based on 
a much poorer observational material (before JD~2\,450\,500). 
An orbital period of about 5600-5700~d is compatible with the data 
pattern in both Fig.~\ref{fig-rzvel-oc2} and Fig.~\ref{fig-rzvel-vgamma} 
but the phase relation between the light-time effect fit to the 
$\Delta (O-C)$ curve and the orbital RV variation phase curve obtained 
with this formal period is not satisfactory.

\subsection{BG~Velorum}
\label{bgvel}

\paragraph*{Accurate value of the pulsation period}
\label{bgvel-period}

The brightness variability of BG~Vel (HD\,78801, $\langle V \rangle
= 7.69$\,mag) was revealed by Cannon \citep{P09}. Much later 
\citet{OL37} independently discovered its light variations but
he also revealed the Cepheid nature and determined the pulsation 
period based on photographic plates obtained at the Riverview 
College Observatory. \citet{vH50} also observed this Cepheid 
photographically in Johannesburg but these early data are 
unavailable, therefore we only mention their studies for historical 
reasons.

This Cepheid is a fundamental-mode pulsator. The O$-$C 
differences of BG~Vel calculated for brightness maxima are 
listed in Table~\ref{tab-bgvel-oc}. These values have been obtained 
by taking into account the constant and linear terms of the 
following weighted parabolic fit:
\vspace{-1mm}
\begin{equation}
C = 2\,453\,031.4706 + 6.923\,843{\times}E + 2.58\times 10^{-8} E^2
\label{bgvel-ephemeris}
\end{equation}
\vspace{-3mm}
$\phantom{mmmmm}\pm0.0020\phantom{}\pm 0.000\,007 \phantom{ml}
\pm 0.27\times 10^{-8}$

\begin{figure}
\includegraphics[height=44mm, angle=0]{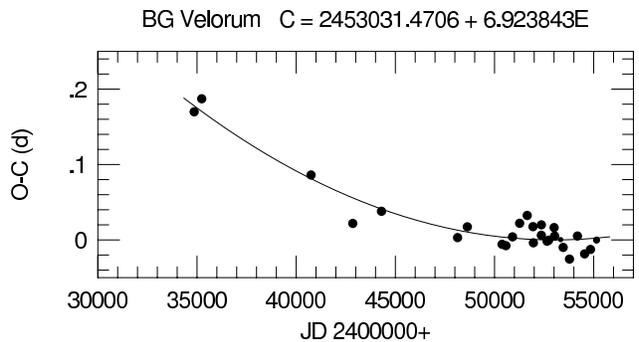}
\caption{O$-$C diagram of BG~Vel. The plot can be
approximated by a parabola indicating a continuously
increasing pulsation period.}
\label{fig-bgvel-oc}
\end{figure}

\noindent The parabolic nature of the O$-$C diagram, i.e., the 
continuous increase in the pulsation period, is clearly seen 
in Fig.~\ref{fig-bgvel-oc}. 
This parabolic trend corresponds to a continuous period increase
of $(5.16 \pm 0.54)\times 10^{-8}$ d\,cycle$^{-1}$, i.e., 
$\Delta P = 0.000272$ d/century. This tiny period increase has 
been also caused by stellar evolution as in the case of RZ~Vel.

\begin{figure}
\includegraphics[height=44mm, angle=0]{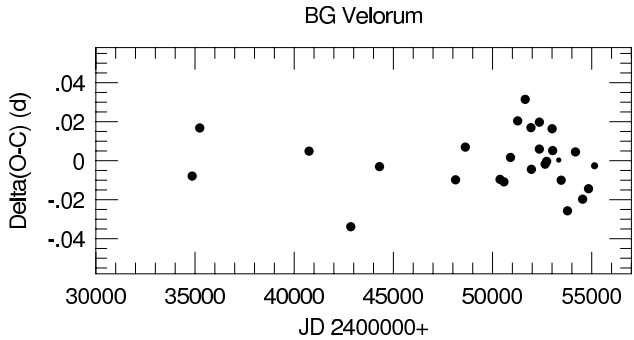}
\caption{$\Delta(O-C)$ diagram of BG~Vel.}
\label{fig-bgvel-oc2}
\end{figure}

The fluctuations around the fitted parabola in
Fig.~\ref{fig-bgvel-oc} do not show any definite pattern: 
see the $\Delta(O-C)$ diagram in Fig.~\ref{fig-bgvel-oc2}.

\begin{table}
\caption{O$-$C values of BG~Vel (description of the 
columns is given in Sect.~\ref{lrtra-period}).}
\begin{tabular}{l@{\hskip2mm}r@{\hskip2mm}r@{\hskip2mm}c@{\hskip2mm}l}
\hline
\noalign{\vskip 0.2mm}
JD$_{\odot}$ & $E\ $ & O$-$C & $W$ & Data source\\
2\,400\,000 + &&&\\
\noalign{\vskip 0.2mm}
\hline
\noalign{\vskip 0.2mm}
34856.5526 & $-$2625 &  0.1699 & 3 & \citet{Wetal58}\\
35237.3813 & $-$2570 &  0.1872 & 3 & \citet{I61}\\
40748.6592 & $-$1774 &  0.0861 & 3 & \citet{P76}\\
42853.4433 & $-$1470 &  0.0219 & 3 & \citet{D77}\\
44300.5426 & $-$1261 &  0.0380 & 3 & \citet{B08}\\
48136.3167 &  $-$707 &  0.0031 & 3 & {\it Hipparcos} \citep{ESA97}\\
48627.9239 &  $-$636 &  0.0174 & 3 & {\it Hipparcos} \citep{ESA97}\\
50379.6329 &  $-$383 & $-$0.0058 & 3 & \citet{B08}\\
50573.4987 &  $-$355 & $-$0.0076 & 3 & \citet{B08}\\
50905.8549 &  $-$307 &  0.0041 & 3 & \citet{B08}\\
51265.9127 &  $-$255 &  0.0221 & 3 & \citet{B08}\\
51646.7345 &  $-$200 &  0.0325 & 3 & \citet{B08}\\
51937.5210 &  $-$158 &  0.0176 & 3 & ASAS \citep{P02}\\
51958.2712 &  $-$155 & $-$0.0038 & 3 & \citet{B08}\\
52359.8640 &   $-$97 &  0.0062 & 3 & ASAS \citep{P02}\\
52359.8778 &   $-$97 &  0.0200 & 3 & \citet{B08}\\
52650.6575 &   $-$55 & $-$0.0017 & 3 & \citet{B08}\\
52726.8212 &   $-$44 & $-$0.0003 & 3 & ASAS \citep{P02}\\
53003.7916 &    $-$4 &  0.0164 & 3 & \citet{B08}\\
53031.4758 &     0   & 0.0052 & 3 & ASAS \citep{P02}\\
53336.1201 &    44   &0.0004 & 1 & {\it INTEGRAL} OMC\\
53460.7390 &    62  & $-$0.0099 & 3 & ASAS \citep{P02}\\
53779.2202 &   108  & $-$0.0254 & 3 & ASAS \citep{P02}\\
54180.8337 &   166  & 0.0052 & 3 & ASAS \citep{P02}\\
54540.8499 &   218  & $-$0.0185 & 3 & ASAS \citep{P02}\\
54838.5810 &   261  & $-$0.0126 & 3 & ASAS \citep{P02}\\
55143.2425 &   305  & $-$0.0002 & 2 & ASAS \citep{P02}\\
\noalign{\vskip 0.2mm}
\hline
\end{tabular}
\label{tab-bgvel-oc}
\end{table}

\paragraph*{Binarity of BG~Vel}
\label{bgvel-bin}

\begin{figure}
\includegraphics[height=49mm, angle=0]{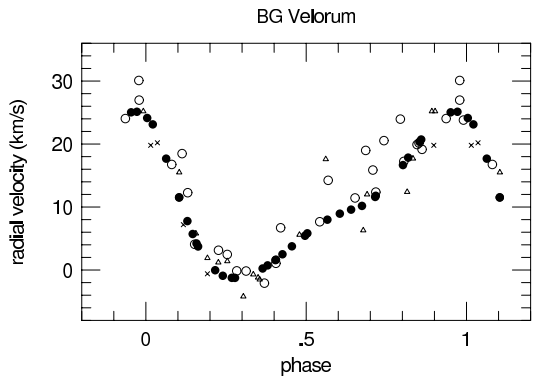}
\caption{Merged RV phase curve of BG~Vel. There is an obvious 
shift between the $\gamma$-velocities valid for the epoch 
of our data obtained in 2005-2006 and 2012-2013 (empty and 
filled circles, respectively). The other symbols are explained
in the text.}
\label{fig-bgvel-vrad}
\end{figure}

\begin{figure}
\includegraphics[height=42mm, angle=0]{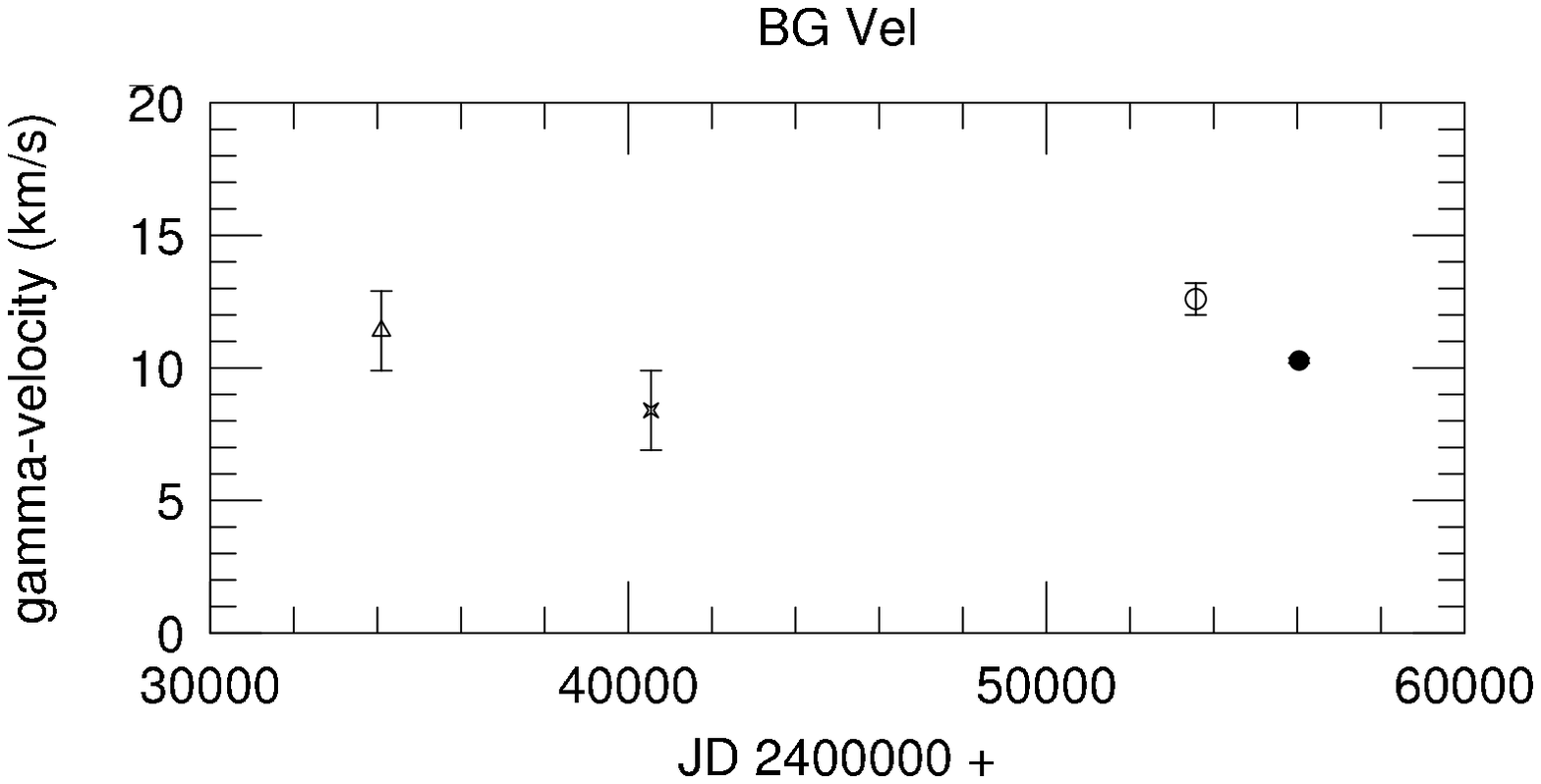}
\caption{$\gamma$-velocities of BG~Vel. The symbols for 
the different data sets are the same as in 
Fig.~\ref{fig-bgvel-vrad}.}
\label{fig-bgvel-vgamma}
\end{figure}

\begin{table}
\caption{RV values of BG Vel from the SSO spectra.
(This is only a portion of the full version available online 
as Supporting Information.)}
\begin{tabular}{lr}
\hline
\noalign{\vskip 0.2mm}
JD$_{\odot}$ & $v_{\rm rad}$  \  \\
2\,400\,000 + &(km\,s$^{-1}$)\\
\noalign{\vskip 0.2mm}
\hline
\noalign{\vskip 0.2mm}
53312.2372 &17.3\\
53364.2219 &$-$0.2\\
53367.1992 &20.5\\
53451.0000 &20.0\\
53452.0021 &23.8\\
\noalign{\vskip 0.2mm}
\hline
\end{tabular}
\label{tab-bgvel-data}
\end{table}

\begin{table}
\caption{CORALIE velocities of BG Vel.
(This is only a portion of the full version 
available online as Supporting Information.)}
\begin{tabular}{lrc}
\hline
\noalign{\vskip 0.2mm}
JD$_{\odot}$ & $v_{\rm rad}$ \ & $\sigma$ \\
2\,400\,000 + &(km\,s$^{-1}$) & (km\,s$^{-1}$)\\
\noalign{\vskip 0.2mm}
\hline
\noalign{\vskip 0.2mm}
55937.7555 & 24.13 & 0.02\\
55938.6241 &  7.77 & 0.02\\  
55939.6522 & $-$1.25 & 0.01\\  
55941.6474 &  7.99 & 0.10\\  
55942.6917 & 11.78 & 0.03\\  
\noalign{\vskip 0.2mm}
\hline
\end{tabular}
\label{tab-bgvel-coralie-data}
\end{table}

There are earlier RV data of this Cepheid obtained by \citet{S55} 
and \citet{LE80}. Variability in the $\gamma$-velocity is seen 
in the merged phase diagram of all RV data of BG~Velorum 
plotted in Fig.~\ref{fig-bgvel-vrad}. In this diagram, our 2005--2006
data (listed in Table~\ref{tab-bgvel-data}) are represented with 
the empty circles, while 2012--2013 data (listed in 
Table~\ref{tab-bgvel-coralie-data}) are denoted by the filled circles, 
the triangles represent Stibbs' data, and the $\times$ symbols refer to 
Lloyd Evans' data. Our RV data have been folded with the period given 
in the ephemeris Equation~(\ref{bgvel-ephemeris}) omitting the quadratic term. 
Data obtained by Stibbs and Lloyd Evans have been phased with the 
same period but a proper correction has been applied to allow 
for the phase shift due to the parabolic O$-$C graph.

The $\gamma$-velocities determined from the individual data sets
are listed in Table~\ref{tab-bgvel-vgamma} and plotted in
Fig.~\ref{fig-bgvel-vgamma}. Since no annual shift is seen 
in the $\gamma$-velocities between two consecutive years (2005--2006
and 2012--2013), the orbital period cannot be short, probably it
exceeds a thousand days.

Similarly to the case of LR~TrA, BG~Vel is also characterized by an 
excessive value for the ratio of RV and photometric amplitudes
indicating the possible presence of a companion 
(see Fig.~\ref{fig-ampratio}).

\begin{table}
\caption{$\gamma$-velocities of BG~Vel.}
\begin{tabular}{lccl}
\hline
\noalign{\vskip 0.2mm}
Mid-JD & $v_{\gamma}$ & $\sigma$ & Data source \\
2\,400\,000+ & (km\,s$^{-1}$)& (km\,s$^{-1}$) & \\
\noalign{\vskip 0.2mm}
\hline
\noalign{\vskip 0.2mm}
34096  &11.4  &1.5& \citet{S55}\\
40545  & 8.4  &1.5& \citet{LE80}\\
53572  &12.6  &0.6& Present paper\\
56043  &10.3  &0.1& Present paper\\
\noalign{\vskip 0.2mm}
\hline
\end{tabular}
\label{tab-bgvel-vgamma}
\end{table}

\section{Conclusions}
\label{concl}

We pointed out that three bright southern Galactic Cepheids,
LR~TrA, RZ~Vel and BG~Vel,
have a variable $\gamma$-velocity implying their membership 
in SB systems. RV values of other target Cepheids observed 
with the same equipment in 2005--2006 and 2012 testify that 
this variability in the $\gamma$-velocity is not of instrumental
origin, nor an artefact caused by the analysis.

The available RV data are insufficient to determine the orbital 
period and other elements of the orbits. However, some inferences
can be made from the temporal variations of the $\gamma$-velocity.
An orbital period of 5600--5700~d of the RZ~Vel system is 
compatible with the data pattern. In the case of BG~Vel, short 
orbital periodicity can be ruled out. For LR~TrA, even the range 
of the possible orbital periods remains uncertain.

The value of the orbital period for SB systems 
involving a Cepheid component is often unknown: according to the 
on-line data base \citep{Sz03a} the orbital period has been 
determined for about 20\% of the known SB Cepheids. The majority 
of known orbital periods exceeds a thousand days.

\begin{figure}
\includegraphics[height=54mm, angle=0]{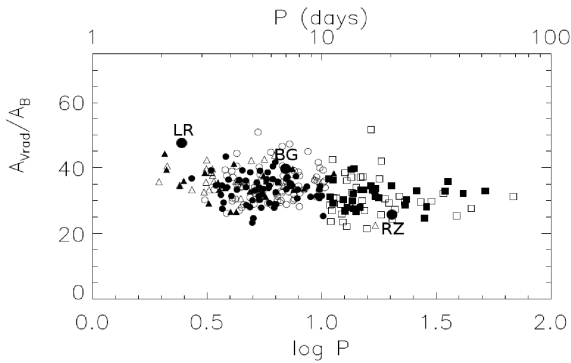}
\caption{The slightly excessive value of the $A_{V_{\rm RAD}}/A_B$
amplitude ratio of LR~TrA and BG~Vel (large circles) with respect 
to the average value characteristic at the given pulsation period
is an independent indication of the presence of a companion star.
This is a modified version of fig.~4f of \citet{KSz09}. The open 
symbols in the original figure correspond to known binaries and 
the filled symbols to Cepheids without known binarity. For the
meaning of various symbols, see \citet{KSz09}.
}
\label{fig-ampratio}
\end{figure}

A companion star may have various effects on the observable
photometric properties of the Cepheid component. Various pieces 
of evidence of duplicity based on the photometric criteria are
discussed by \citet{Sz03b} and \citet{KSz09}. As to our
targets, there is no obvious sign of a companion from optical
multicolour photometry. This indicates that the companion star 
cannot be much hotter than any of the Cepheids discussed here. 
There is, however, a phenomenological parameter, viz. the ratio 
of RV to photometric amplitudes \citep{KSz09} whose excessive
value is a further hint at the probable existence of a
companion for both LR~TrA and BG~Vel (see Fig.~\ref{fig-ampratio}).
Moreover, the {\it IUE} spectra of bright Cepheids 
analysed by \citet{E92} gave a constraint on the temperature 
of a companion to remain undetected in the ultraviolet spectra: 
in the case of RZ~Vel, the spectral type of the companion cannot 
be earlier than A7, while for BG~Vel this limiting spectral type 
is A0. Further spectroscopic observations are necessary to 
characterize these newly detected SB systems.

Our findings confirm the previous statement by \citet{Sz03a} 
about the high percentage of binaries among classical Cepheids 
and the observational selection effect hindering the discovery 
of new cases (see also Fig.~\ref{fig-comparison}).

Regular monitoring of the RVs of a large
number of Cepheids will be instrumental in finding 
more SBs among Cepheids. RV data to be obtained with the 
{\it Gaia} astrometric space probe (expected launch: 2013 
September) will certainly result in revealing new SBs among 
Cepheids brighter than the 13--14th magnitude \citep{Eyetal12}.
In this manner, the `missing' SBs among Cepheids inferred
from Fig.~\ref{fig-comparison} can be successfully revealed
within few years.

\section*{Acknowledgments} 

This project has been supported by the 
ESTEC Contract No.\,4000106398/12/NL/KML, the Hungarian OTKA 
Grants K76816, K83790, K104607, and MB08C 81013, as well as the 
European Community's Seventh Framework Program (FP7/2007-2013) 
under grant agreement no.\,269194,  and the ``Lend\"ulet-2009'' 
Young Researchers Program of the Hungarian Academy of Sciences. 
AD was supported by the Hungarian E\"otv\"os Fellowship. 
AD has also been supported by a J\'anos Bolyai Research Scholarship 
of the Hungarian Academy of Sciences. AD is very thankful 
to the staff at The Lodge in the Siding Spring Observatory 
for their hospitality and very nice food, making the 
time spent there lovely and special.
Part of the research leading to these results has received 
funding from the European Research Council under the European 
Community's Seventh Framework Programme (FP7/2007--2013)/ERC grant 
agreement no.\,227224 (PROSPERITY).
The {\it INTEGRAL\/} photometric data, pre-processed by 
ISDC, have been retrieved from the OMC Archive at CAB (INTA-CSIC). 
We are indebted to Stanley Walker for sending us some
unpublished photoelectric observational data. Our thanks are 
also due to the referee and Dr. M\'aria Kun for their critical 
remarks leading to a considerable improvement in the presentation 
of the results.

\bsp

\label{lastpage}

\end{document}